\documentclass[twocolumn,9pt]{extarticle}
\RequirePackage{lmodern}
\usepackage[T1]{fontenc}
\usepackage{xcolor}
\usepackage{graphicx}

\usepackage{enumitem}
\usepackage{amsfonts}

\usepackage{dcolumn}
\usepackage{bm}
\usepackage[mathlines]{lineno}
\usepackage[lmargin=48.5pt, rmargin=53.5pt, tmargin=40pt, bmargin=60pt, columnsep=11.5pt]{geometry}
\usepackage{caption}
\usepackage{titling}
\usepackage[superscript, biblabel]{cite}

\usepackage{floatrow}
\usepackage[normalem]{ulem}

\usepackage{physics}
\usepackage{setspace}
\usepackage{subfiles} 
\usepackage{comment}
\usepackage[euler]{textgreek}

\pretitle{\begin{flushleft}}
\posttitle{\end{flushleft}}
\preauthor{\begin{flushleft}}
\postauthor{\end{flushleft}}

\makeatletter

\newcommand\blfootnote[1]{%
  \begingroup
  \renewcommand\thefootnote{}\footnote{#1}%
  \addtocounter{footnote}{-1}%
  \endgroup
}

\begin{document}

\title{\textsf{\textbf{{\fontsize{25}{60}\selectfont Spectrally reconfigurable quantum emitters enabled by optimized fast modulation}}}}


\author{Daniil M. Lukin,$^{\dagger1}$  Alexander D. White,$^{\dagger1}$ Rahul Trivedi,$^{1}$ Melissa A. Guidry,$^{1}$ Naoya Morioka,$^2$ Charles Babin,$^2$ \"{O}ney O. Soykal,$^3$ Jawad Ul-Hassan,$^4$ Nguyen Tien Son,$^4$ Takeshi
Ohshima,$^5$ Praful K. Vasireddy,$^6$ Mamdouh H.  Nasr,$^6$ Shuo Sun,$^1$ Jean-Philippe W. MacLean,$^1$ Constantin Dory,$^1$ Emilio A. Nanni,$^6$ J\"org Wrachtrup,$^2$ Florian Kaiser,$^2$ and Jelena Vu{\v c}kovi\'c$^1$}

\date{\vspace{-4ex}}

\begingroup
\let\center\flushleft
\let\endcenter\endflushleft
\maketitle
\endgroup

\noindent\textsf{\textbf{The ability to shape photon emission facilitates strong photon-mediated interactions between disparate physical systems, thereby enabling applications in quantum information processing,  simulation and communication.
Spectral control in solid state platforms such as color centers, rare earth ions, and quantum dots is particularly attractive for realizing such applications on-chip. Here we propose the use of frequency-modulated optical transitions for spectral engineering of single photon emission. 
Using a scattering-matrix formalism, we find that a two-level system, when modulated faster than its optical lifetime, can be treated as a single-photon source with a widely reconfigurable photon spectrum that is amenable to standard numerical optimization techniques. To enable the experimental demonstration of this spectral control scheme, we investigate the Stark tuning properties of the silicon vacancy in silicon carbide, a color center with promise for optical quantum information processing technologies. We find that the silicon vacancy possesses excellent spectral stability and tuning characteristics, allowing us to probe its fast modulation regime, observe the theoretically-predicted two-photon correlations, and demonstrate spectral engineering.
Our results suggest that frequency modulation is a powerful technique for the generation of new light states with unprecedented control over the spectral and temporal properties of single photons.
}}
\blfootnote{
\begin{itemize}[leftmargin=*]
 \item[$\dagger$] These authors contributed equally 
 \item[1] E. L. Ginzton Laboratory, Stanford University, Stanford, California 94305, USA
 \item[2] 3rd Institute of Physics, University of Stuttgart and Institute for Quantum Science and Technology IQST, 70569, Stuttgart, Germany
 \item[3] Booz Allen Hamilton, McLean, VA, 22102 USA
 \item[4] Department of Physics, Chemistry and Biology, Link\"oping University, SE-58183, Link\"oping, Sweden
 \item[5] National Institutes for Quantum and Radiological Science and Technology, Takasaki, Gunma 370- 1292, Japan
 \item[6] SLAC National Accelerator Laboratory, Stanford University, 2575 Sand Hill Road, Menlo Park, California 94025, USA
\end{itemize}
}

Photon-mediated interactions between quantum systems are at the heart of a number of quantum information applications including quantum networking, simulation and computation\cite{o2009photonic, Reiserer2015,  hensen2015loophole, QuantumNetworkSiV2019, gonzalez2015subwavelength, douglas2015quantum, arguello2019analogue}. Because the physical characteristics of nodes can differ drastically, the ability to spectrally control photon emission enables networks composed of disparate physical systems. Spectral-shaping techniques can enable the scaling of quantum simulators despite the inhomogeneities in the comprising qubits\cite{altman2019quantum, fotso2016suppressing}.
Spectral shaping is also necessary to maximize the absorption fidelity of a photon by an atom\cite{stobinska2009perfect}, and to maximize photon-photon interference visibility in the presence of imperfections\cite{rohde2005optimal}. 
Moreover, frequency-encoded quantum states can be used for high-dimensional entanglement protocols\cite{Lukens2017,kues2017chip}.

The prevalent approach for deterministic generation of single photons is spontaneous emission from a two‐level system (TLS)\cite{JinQDPhotonSource2019}. However,  since  unperturbed  spontaneous  emission only produces photons with a Lorentzian spectrum, significant effort has been devoted to exploring more complex systems for spectral control of single photon emission. For instance, a TLS with a time-dependent coupling to a cavity has been studied for symmetrization of single-photon wavepackets\cite{UltrafastEmissionJin, pagliano2014dynamically}, and cascaded three-level systems and lambda systems have been used for partly-stimulated two-photon emission\cite{breddermann2016tailoring} and Raman emission\cite{keller2004continuous, pursley2018picosecond}, respectively. A system-agnostic approach to photon shaping is post-processing emitted photons using cavities\cite{srivathsan2014reversing}, electro-optic phase modulators\cite{specht2009phase} and nonlinear frequency conversion\cite{matsuda2016deterministic, lavoie2013spectral}, at the expense of additional system complexity and loss. 

In this Article, we propose a comprehensive alternative to these techniques that requires only a TLS whose transition energy can be rapidly modulated.
The time-modulated TLS has been studied in other contexts since the early days of quantum mechanics\cite{silveri2017quantum}; its application in atomic systems\cite{baruch1992ramsey}, superconducting qubits\cite{li2013motional}, gate-defined quantum dots\cite{koski2018floquet}, and solid state defects \cite{metcalfe2010resolved, forster2015landau, MiaoElectricallyCarbide, chen2018orbital, schadler2019electrical} has allowed for the demonstration of fundamental phenomena such as spectral sideband formation, Landau-Zener-St\"{u}ckelberg interference, and motional averaging. Here, we examine the fast time-modulated TLS as a single-photon source. To this end, we study the few-photon scattering properties of the modulated TLS, as well as its single-photon-emission fidelity under pulsed resonant drive. We find that, remarkably, in the fast modulation regime (\textit{i.e.}, modulation faster than the optical lifetime) the modulated TLS can be treated like a conventional two-level system but with an exotic, reconfigurable spectrum. We experimentally characterize static and time-modulated Stark shift in the negatively-charged single silicon vacancy (V$_{\text{Si}}^-$) color centers in 4H silicon carbide (4H-SiC), and observe that the optical coherence properties of the V$_{\text{Si}}^-$ are preserved even under high-amplitude modulation. With this system, we investigate the few-photon scattering from a modulated quantum emitter, and demonstrate the proposed spectral engineering of single photons from a solid-state TLS. Finally, we demonstrate pulsed optical orbital control under modulation through measurement of Rabi oscillations and Ramsey interference.

\section*{{\fontsize{11}{11}\textsf{\textbf{Continuous-wave scattering off a modulated \\ two-level system\vspace{-1.2 ex}}}}}

\begin{figure*}[h!]
\centering
\floatbox[{\capbeside\thisfloatsetup{capbesideposition={right,top},capbesidewidth=4.5cm}}]{figure}[11.2cm]
{\caption{\vspace{1pt}\\
\textbf{Figure 1 \textbar \hspace{0.1pt} Theoretical analysis of photon scattering from a modulated TLS}. \textbf{a} Schematic depiction of single-photon scattering from a modulated TLS with $\Delta(t) = A \sin(\Omega t)$. A continuous-wave photon in the input optical channel is scattered into an output optical channel to produce a superposition of continuous-wave single-photons at frequencies $\nu + p\Omega \ \forall \ p\in\mathbb{Z}$. \textbf{b} The spectrum of the single-photon state in the output channel as a function of frequency $\omega$ on exciting a modulated two-level system ($\Omega = 2.5\gamma$) with a Gaussian single-photon wavepacket centered at $\nu$ with FWHM $0.625\gamma$. The color scale shows the magnitude squared of the output spectrum. \textbf{c} The total transmission in the output optical channel as a function of the frequency of the input photon. \textbf{d} The amplitude squared of the output two-photon wavepacket as a function of the initial time-instant $t$ and the time-difference $\tau$. \textbf{e} The two-photon correlation function as a function of the time-difference between the two photons in the output wavefunction. The dashed line indicates $g^{(2)}(\tau)$ for an unmodulated TLS. $A = 5\gamma$ is used in all simulations. \label{fig:scattering_theory}}}
{ \includegraphics[width=11cm]{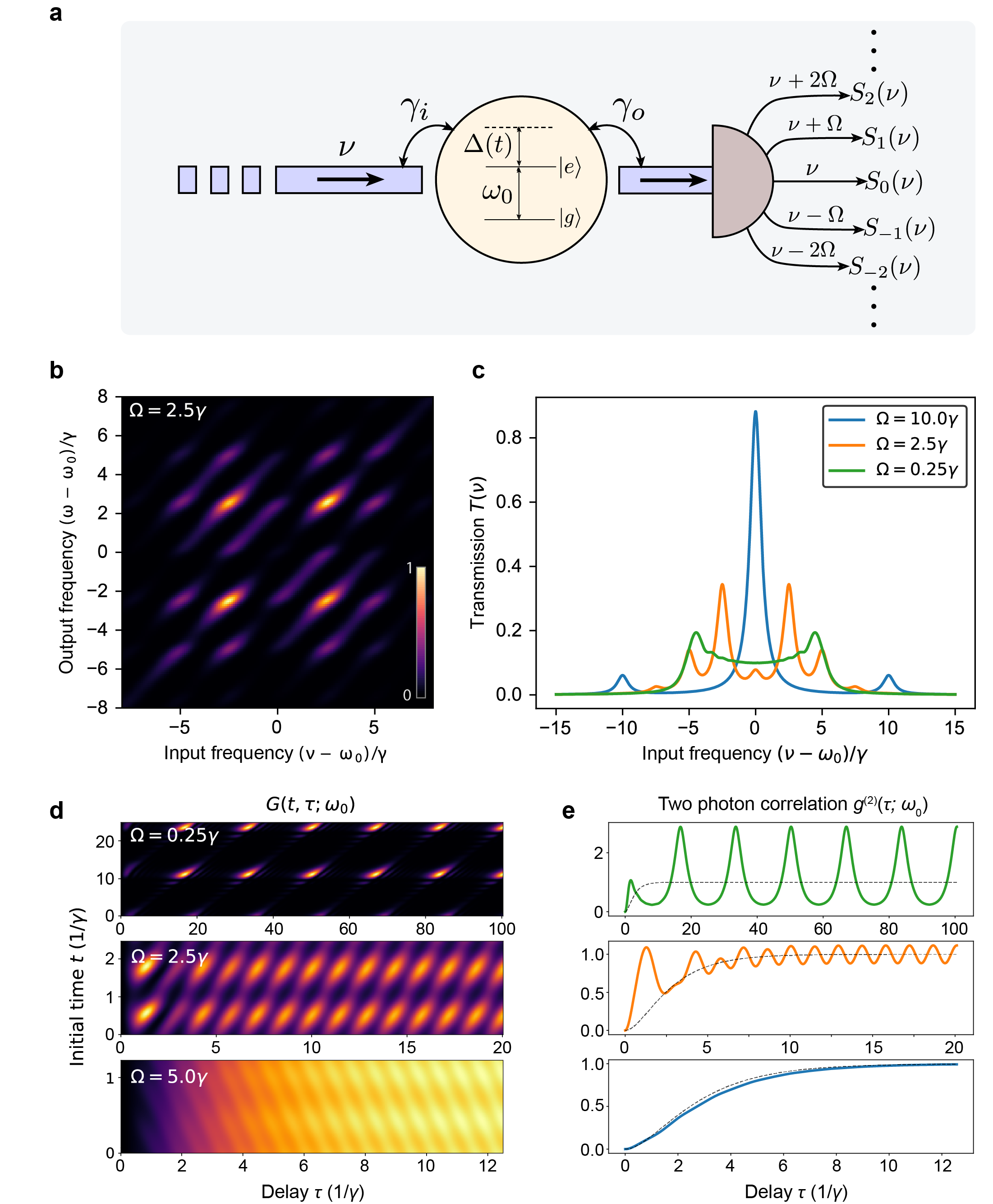}}
\end{figure*}

We study the modulated TLS via the quantum optics formalism of Markovian open quantum systems, with the goal of understanding the few-photon statistics and  the single-photon spectra of scattered photons. We consider a modulated TLS driven by an arbitrary periodic modulation $\Delta(t)$ with period~$2\pi/\Omega$,
\begin{align}
    H_\text{sys}(t) = (\omega_0 + \Delta(t)) \sigma^\dagger \sigma
\end{align}
where $\omega_0$ is the resonant frequency of the TLS and $\sigma$ is the de-excitation operator. Furthermore, we impose a decay rate $\gamma$ corresponding to the system lifetime. The Floquet eigenstates for this Hamiltonian can be computed analytically (see Supplementary Information).
To study the form of the emitted photon wavepacket under excitation by a weak coherent state, we use single- and two-photon scattering matrices of the modulated TLS. While scattering matrices are traditionally computed only for time-independent systems, it was recently shown that they can be defined and computed for time-dependent systems\cite{trivedi2018scattering, Trivedi2020Analytic}. As is shown in the Supplementary Information, the single-photon scattering matrix through the modulated TLS $S(\omega, \nu)$, defined as the amplitude of producing an output photon at frequency $\omega$ when the system is excited with an input photon at frequency $\nu$, can be expressed as:
\begin{align}
\begin{split}
    &S(\omega, \nu) = \sum_{p=-\infty}^\infty S_p(\nu) \delta(\omega - \nu - p\Omega) \text{,} \\
     &\text{where } \ S_p(\nu) = -\sum_{m=-\infty}^\infty  \frac{\sqrt{\gamma_i \gamma_o}\alpha_m^* \alpha_{m + p}}{\gamma / 2 + \textrm{i}(\omega_0 + m\Omega - \nu )} 
\end{split}
\end{align}
\begin{figure}[h!]
{\includegraphics[width = 8.4cm]{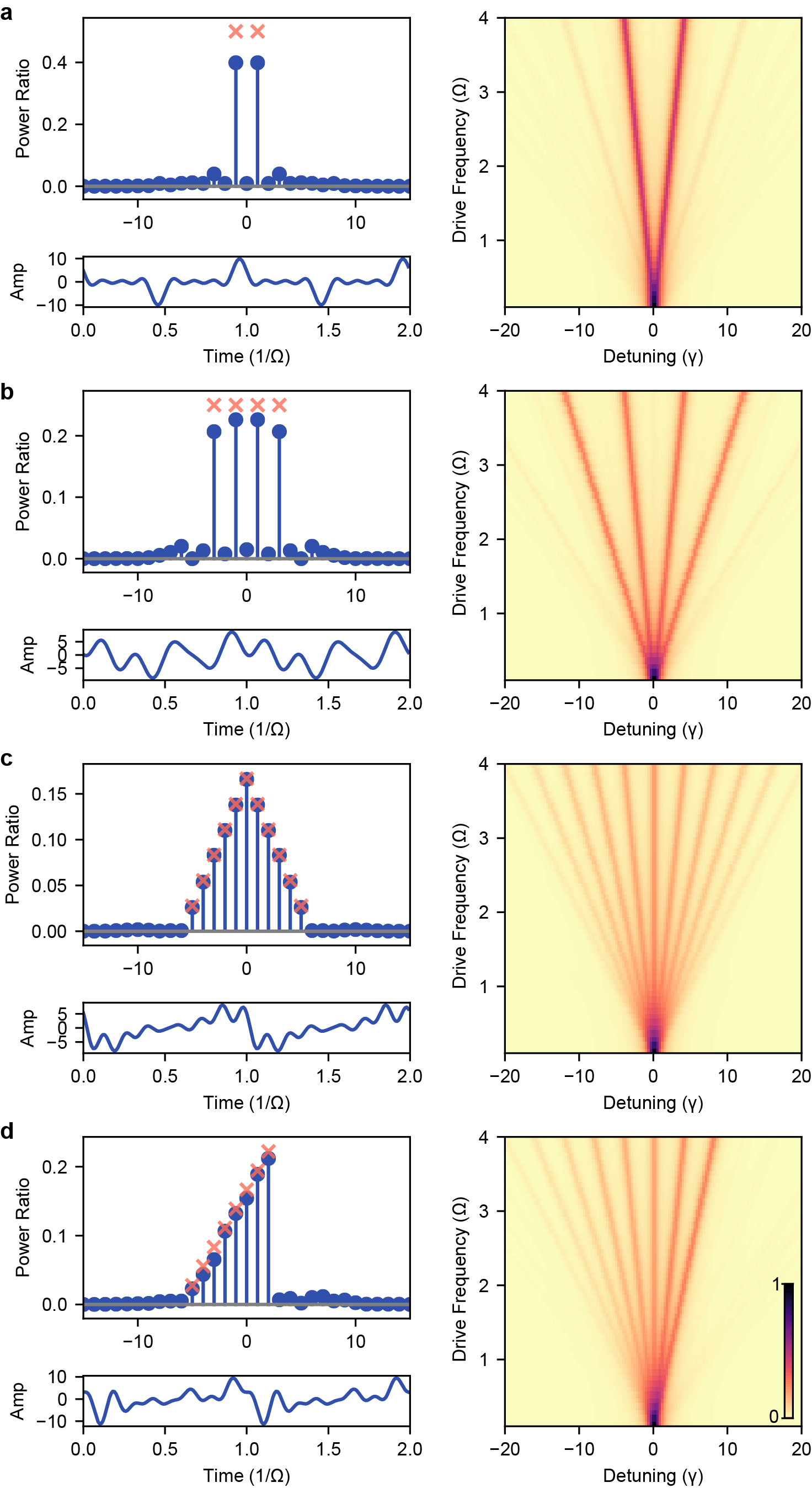}}
{\caption{\textbf{Figure 2 \textbar \hspace{0.1pt} Spectral optimization of Floquet states.} \textbf{a-d.} Four examples of spectral optimization are shown. Left panels: the target spectrum is denoted by red crosses (omitted for target value of zero). The optimized spectrum is shown in blue. Below, the corresponding $\Delta(t)$ is shown for two periods. Right panels: The resulting Floquet spectra are shown, with optical detuning on the x-axis and fundamental drive frequency $\Omega$ on the y-axis. The color scale gives the relative magnitude squared of the Floquet spectrum. The spectral harmonics of a Floquet eigenstate are dictated by the normalized drive amplitude A/$\Omega$. By scaling $\Delta(t)$ with $\Omega$, the separation of spectral peaks can be controlled while retaining their amplitude.} \label{fig:optimized_spectra}}
\end{figure} 
\hspace{-7pt}
where $\gamma_{i,o}$ is the coupling rate into the input (output) channel, $\gamma = \gamma_i + \gamma_o$ is the total decay rate of the two-level system, and  $\alpha_m$ is the Fourier-series coefficient of the phase $\exp(-\textrm{i}\int_0^t \Delta(t')dt')$ accumulated by the excited state of the modulated TLS up to time $t$, corresponding to the harmonic $\exp(-\textrm{i}m\Omega t)$.
We note that the form of $S(\omega, \nu)$ implies that the output photon state corresponding to an input photon state at frequency $\nu$ has frequencies $\nu + p\Omega, \ p\in \mathbb{Z}$ as a consequence of the periodic modulation of the emitter frequency (see Fig.~1a). 
In particular, a narrowband incident photon wavepacket would scatter from the modulated two-level system into different modulation sidebands (Fig.~\ref{fig:scattering_theory}b). 
In the slow modulation regime ($\Omega_0 \ll \gamma$), the total transmission from the TLS ($T(\nu)$), computed as a sum of transmission into individual sidebands ($|S_p(\nu)|^2$) , is simply a time-average of transmission spectra with different resonant frequencies. 
The fast modulation regime ($\Omega_0 \gg \gamma$) is characterized by distinctive sidebands with high transmission at the resonant frequency (see Fig.~\ref{fig:scattering_theory}c).  
The scattering amplitudes into different sidebands can be controlled by an appropriate choice of the time-dependent frequency modulation $\Delta(t)$.

\begin{figure*}[h!]
\centering
\includegraphics[width = 15.73cm]{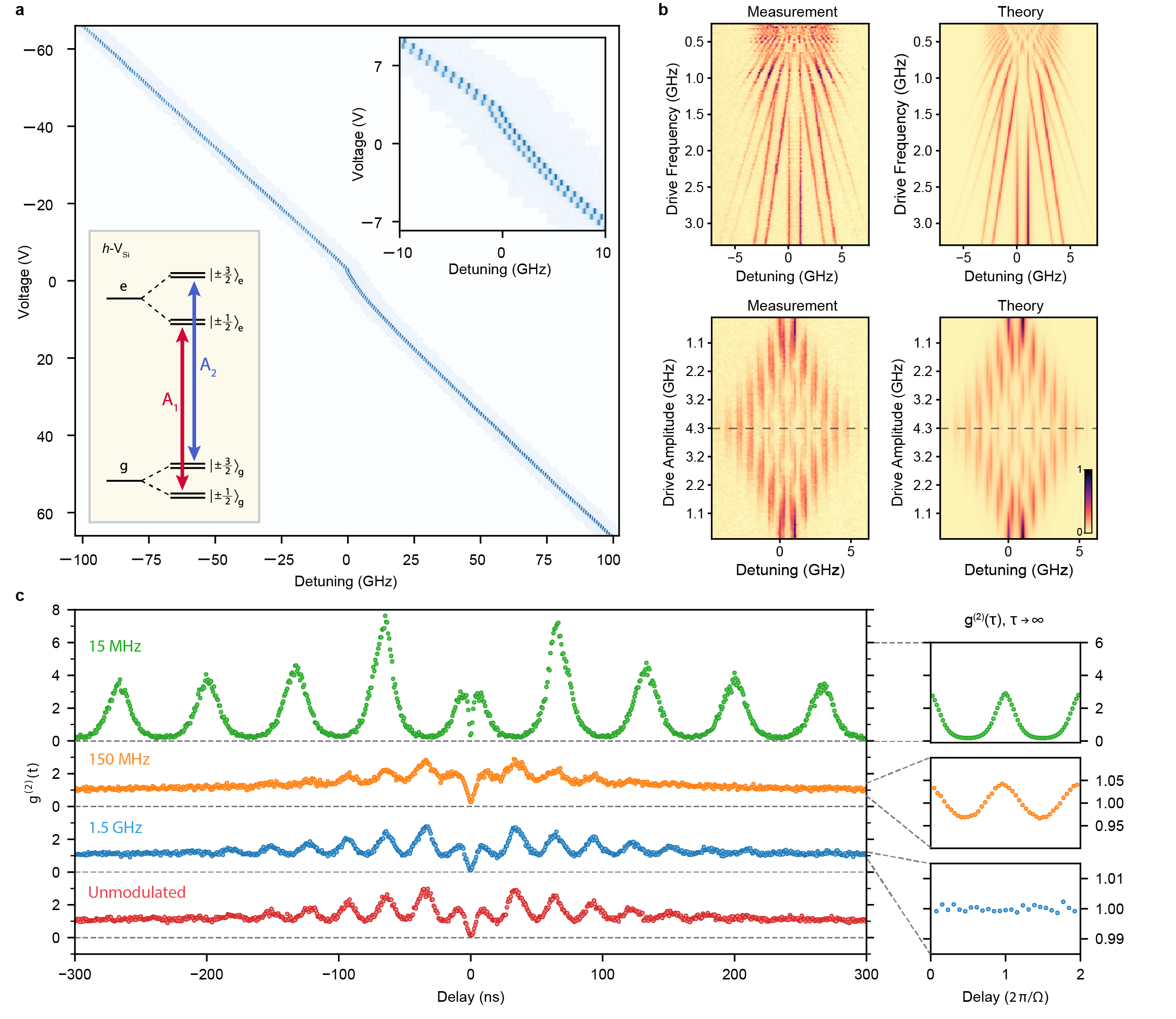}
\caption{\textbf{Figure 3 \textbar \hspace{0.1pt} Few-photon scattering off a single Stark-modulated V\textsubscript{Si}.} \textbf{a} Static tuning properties of the \textit{h}-V\textsubscript{Si}, showing a wide tuning range of $200$~GHz. Left inset shows the level structure of the two optical transitions. Right inset shows a close-up of a nonlinear behavior likely due to field rectification by charge traps\cite{ElectricalTuningNVBassett2011}. \textbf{b} Spectral signatures of Floquet states in the \textit{k}-V\textsubscript{Si} for $\Delta(t) = A\sin{(\Omega t)}$ harmonic drive, for swept $\Omega$ under a fixed amplitude of $A=3$~GHz (upper), and swept $A$ with fixed $\Omega/2\pi= 750$~MHz
(lower). Color corresponds to the normalized photon counts emitted into the phonon sideband.
\textbf{c} Measurement of the second-order photon correlation --- $g^{(2)}(\tau)$ --- under weak coherent excitation in the slow (15~MHz), intermediate (150~MHz) and fast (1.5~GHz) modulation regimes. The modulation-independent oscillations at short time delays originate from the interference of the multiple states in the ground manifold, as discussed in the Supplementary Information. In the limit of long time delay, $g^{(2)}(\tau)$ of a modulated emitter becomes periodic. To resolve the fine oscillatory features, we average the $g^{(2)}(\tau)$ data over many microwave periods  (up to $\tau=200\mu$s), shown in the right panel. }
\label{fig:dcac}
\end{figure*}

For non-classical light generation, especially single-photon generation, it is of utmost importance to understand the statistics of photons transmitted through the modulated TLS. In this context, of particular interest is the scattering of a monochromatic two-photon pair at frequency $\nu$ from the modulated TLS. The output state $\ket{\psi_\text{out}(\nu)}$ can be described by its wavefunction $\psi_\text{out}(t, t + \tau; \nu)$, which represents the amplitude of one photon in the output being emitted at time $t$ and the second photon being emitted after a delay of $\tau$. A detailed calculation of $\psi_\text{out}(t, t+ \tau; \nu)$ in terms of the modulation $\Delta(t)$ within the framework of the scattering theory is outlined in the Supplementary Information. Figure~\ref{fig:scattering_theory}d shows $G(t, \tau; \omega_0) = |\psi_\text{out}(t, t + \tau; \omega_0)|^2$ under different modulation regimes; in all cases, it can be seen that $G(t, 0; \nu) = 0$. This implies that any modulated TLS shows perfect photon antibunching. In the slow modulation regime, the amplitude of the two-photon wavepacket varies periodically with the delay $\tau$ as a consequence of the periodic drive. In the fast modulation regime, a continuous-wave excitation only significantly addresses one of the Floquet sidebands and consequently the time-domain scattering properties resemble those of an unmodulated TLS. Figure~\ref{fig:scattering_theory}e shows the two-photon correlation function $g^{(2)}(\tau; \omega_0) = \langle G(t, \tau; \omega_0) \rangle_t / T^2(\omega_0)$. As with $G(t, \tau; \omega_0)$, we observe that $g^{(2)}(\tau; \omega_0)$ oscillates with $\tau$ in the slow modulation regime, and resembles the correlation function of an unmodulated TLS in the fast modulation regime.

Using a scattering matrix formalism, we have found that in the limit of fast drive a modulated TLS behaves very much like a conventional single-photon source but with a controllable spectrum, determined by the form of the drive $\Delta(t)$. Thus, a modulated TLS can serve as a source of photons whose spectral composition is amenable to standard numerical optimization techniques; one can optimize $\Delta(t)$ given a target single-photon spectrum. Figure~\ref{fig:optimized_spectra} shows examples of single-photon spectra that can be obtained via such optimization (see Methods). We note that due to the periodicity of $\Delta(t)$, this approach is limited to producing photons with discrete, comb-like spectra. This restriction is lifted for aperiodic $\Delta(t)$, which we discuss further in the Supplementary Information.

\section*{{\fontsize{11}{11}\textsf{\textbf{Reconfiguring photon spectra using optimized \\ periodic modulation\vspace{-1.2 ex}}}}}


To realize the proposal experimentally, it is essential to identify a single-photon source amenable to fast, broadband modulation. A suitable solid-state defect must possess a widely tunable optical transition while displaying minimal spectral diffusion in a stationary state and under rapid modulation. Here, we study the Stark tuning and the spectral stability of a single V$_{\text{Si}}^-$ color center in 4H-SiC (abbreviated V\textsubscript{Si} henceforth), and find the V\textsubscript{Si} to satisfy these requirements. The V\textsubscript{Si} is a color center with promise for quantum information processing technologies due to its long spin coherence time\cite{Widmann2015CoherentTemperature}, unique 3/2 spin system\cite{SoykalSiliconVacancy, Morioka2020SpinControlled}, and compatibility with scalable photonic architectures\cite{4HSiCPhotonics2019}. The V\textsubscript{Si} spectrum comprises two optical transitions, which are separated by 1~GHz via spin-spin coupling (corresponding to spin $\pm1/2$ and $\pm3/2$ sublevels). The V\textsubscript{Si} occurs at two inequivalent lattice sites in the 4H-SiC lattice, denoted by \textit{h} and \textit{k}, with the zero-phonon lines at 861~nm and 916~nm, respectively. The properties of the defects are largely similar\cite{Banks2019Resonant4H-SiC, Nagy2018High-fidelityCarbide}; one distinction is that the optical coherence of \textit{k}-V\textsubscript{Si} is more robust against dephasing caused by acoustic phonons (characterized by narrower linewidths at elevated temperatures)\cite{Udvarhelyi2020Vibronic}.

We first probe the static Stark shift of the V\textsubscript{Si} by applying a voltage across gold electrodes fabricated on the \textit{a}-cut surface of 4H-SiC, oriented to apply the field along the axis of symmetry of the defect (see Supplementary Information), and measure the single-defect spectrum (see Methods). All measurements are performed at $5$~K via resonant absorption spectroscopy, \textit{i.e.}, photoluminescence excitation (PLE). As shown in Fig.~\ref{fig:dcac}a, we observe that the zero-phonon line of the V\textsubscript{Si} can be tuned by $200$~GHz without degradation of spectral properties: in other solid-state emitters, the degradation manifests as blinking, charge conversion, or carrier tunneling\cite{VVStarkdelasCasas2017, aghaeimeibodi2019large} The V\textsubscript{Si} does require the periodic application of an above-resonant laser for charge stabilization\cite{Banks2019Resonant4H-SiC, Nagy2018High-fidelityCarbide}, but we do not observe an increased rate of charge conversion with the application of a bias voltage. The static Stark shift measurement is performed by incrementing the applied voltage in steps of $0.5$~V and sweeping a tunable laser, programmed to track the frequency of the V\textsubscript{Si} as it shifts.  From electrostatic simulation of the electrodes and the Lorentz local field approximation\cite{PhysRevLett.97.083002}, we calculate the local electric field strength at the V\textsubscript{Si} location (assuming the defect position is accurate within 1~\textmugreek m\textsuperscript{3}), and deduce a strong Stark shift of $3.65\pm0.09$~GHz/(MV/m). This corresponds to an electric dipole moment of $0.72\pm0.02$~Debye, in disagreement with the theoretical prediction of 0.2~Debye\cite{Udvarhelyi2019InversionSymm}. We note that a recent experimental study in V\textsubscript{Si} ensembles estimated the dipole moment to be 0.18~Debye\cite{StarkCarbideRuhl2020}. Nevertheless, the wide-range, high resolution characterization of the Stark shift of single color centers presented in this work gives us confidence in the dipole moment magnitude we report. A further investigation of spectral diffusion properties and Stark shift for \textit{k}- and \textit{h}-V\textsubscript{Si} is presented in the Supplementary Information.

\begin{figure*}[ht!]
\centering
\includegraphics[width=150mm]{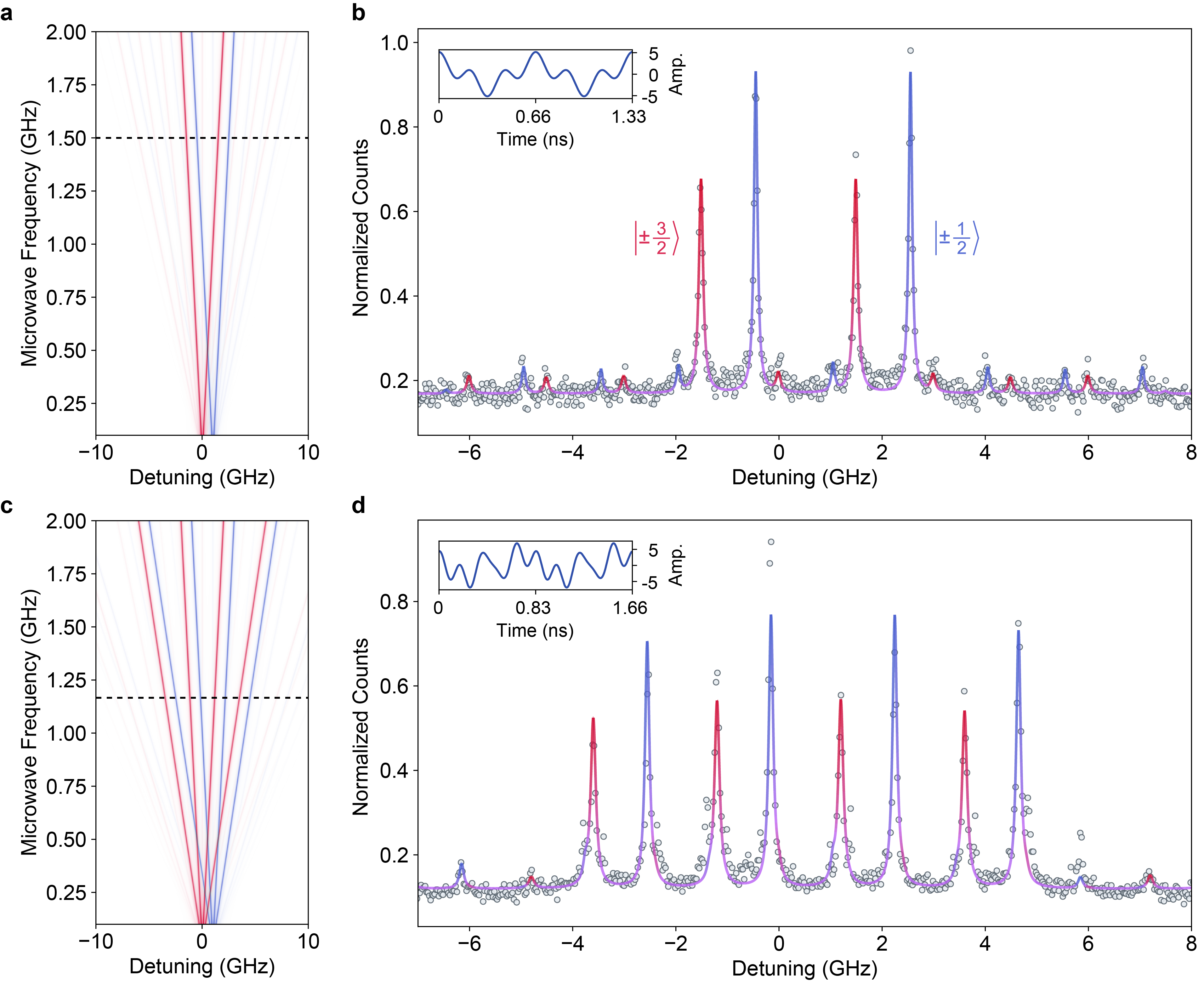}
\caption{\textbf{Figure 4 \textbar \hspace{0.1pt} Spectral optimization of the V\textsubscript{Si}}. \textbf{a} Theoretical spectrum under 2-color optimized microwave drive, where the red (blue) represents the $\pm3/2$ ($\pm1/2$) transitions. Dashed line indicates the slice that is produced experimentally. \textbf{b} The experimentally measured spectrum. The datapoints are overlaid with the theoretical plot of the target Floquet state. The color of the fit line corresponds to the contributing transition, where red (blue) represents the $\pm3/2$ ($\pm1/2$) transition. Inset: Two periods of the time domain signal. The x-axis is time (ns) and the y-axis is modulation amplitude (GHz). \textbf{c-d} correspond one-to-one with \textit{a-b} for a four-color superposition. The \textit{k}-V\textsubscript{Si} is used for this demonstration.} 
\label{fig:multi_color_photons}
\end{figure*}

We then proceed to characterize the V\textsubscript{Si} under Stark modulation, in order to observe spectra of Floquet eigenstates which have been previously seen in other solid state quantum emitters such as the NV and SiV centers in diamond \cite{chen2018orbital, maity2020coherent}, the divacancy in SiC\cite{MiaoElectricallyCarbide}, and quantum dots\cite{metcalfe2010resolved}. Applying sinusoidal modulation $\Delta(t)$ for a range of frequencies and amplitudes, we observe that the V\textsubscript{Si} spectrum matches the prediction of the scattering matrix theory (Fig.~\ref{fig:dcac}b). Crucially, we see that the V\textsubscript{Si} spectral stability is not impacted by the periodic drive. A detailed analysis of the spectral stability of the V\textsubscript{Si} under modulation is presented in the Supplementary Information. We then study the two-photon scattering properties of the V\textsubscript{Si}. As expected for a single-photon emitter, we observe antibunching for all modulation frequencies (Fig.~\ref{fig:dcac}c). Additionally, we observe two independent effects: 1) the oscillations in $g^{(2)}(\tau)$ due to the emitter modulation, present for all $\tau$; and 2) a modulation-independent signature of interference between the four ground states in the ground manifold, decaying exponentially with $\tau$ due to the spin mixing via the intersystem crossing. A detailed analysis of this interference effect is presented in the Supplementary Information. The 15~MHz, 150~MHz, and 1.5~GHz modulation frequencies probe the slow, intermediate, and fast regimes, respectively.
In the slow regime, the shape of $g^{(2)}(\tau)$ is strongly modified by the Stark modulation. As the modulation frequency is increased, the $g^{(2)}(\tau)$ becomes practically indistinguishable from that of the unmodulated emitter, in agreement with the scattering-matrix theory.

To generate spectrally-engineered Floquet states, we drive the V\textsubscript{Si} periodically with optimized harmonics. To demonstrate the range of achievable spectra, we prepare the V\textsubscript{Si} in Floquet states that emit photons in an equal superposition of two and four colors. We restrict the optimization to a bandwidth of $6$~GHz, limited by the external microwave losses. Figure~\ref{fig:multi_color_photons}a(c) shows the theoretical spectrum resulting from the optimization of a two- (four-) color photon spectrum. The corresponding microwave drives $\Delta(t)$ are shown in the Fig.~\ref{fig:multi_color_photons}b(d) insets, and more detail is given in the Supplementary Information. The experimentally-generated two- (four-) color state is shown in Fig.~\ref{fig:multi_color_photons}b(d). We note that the PLE measurement is performed in the weak-excitation regime, and thus probes the spectral composition of the photons spontaneously emitted by the system. \cite{trivedi2019photon}

\begin{figure*}[h!]
\centering
\floatbox[{\capbeside\thisfloatsetup{capbesideposition={right,top},capbesidewidth=4.5cm}}]{figure}[12cm]
{\caption{\textbf{\vspace{1pt}\\
Figure 5 \textbar \hspace{0.1pt} Fidelity of single-photon emission for pulsed excitation of a modulated two-level system.} \textbf{a.} We analyze numerically a V\textsubscript{Si} modulated with $\Delta(t) = (16\mbox{ GHz})\sin(2\pi t(10\mbox{ GHz}))$ and excited by a Gaussian laser pulse (green) centered at the first sideband. \textbf{b}. Numerical simulation of the population dynamics of a two-level system (initially in the ground state) shows a strong dependence on the relative phase between the microwave drive and the optical pulse (FWHM of 166~ps, pulse area of 10$\pi$). \textbf{c} Expected number of photons emitted for different pulse widths. The deviation from the unmodulated emitter at a pulse length of $10^{-2}/\gamma$ corresponds to the intermediate pulse-length regime where the optical pulse is broad enough to couple to multiple sidebands but is not yet broad enough to cover the entire spectrum. \textbf{d} Pulse-wise two-photon correlation function for different pulse widths, indicating the contribution of multi-photon emission events. $g^{(2)}[0]$ approaching zero with the expected photon number remaining near unity indicates a single-photon source that emits one-and-only-one photon with high fidelity.
\label{fig:pi_pulse_drive} }}
{\includegraphics[width=12cm]{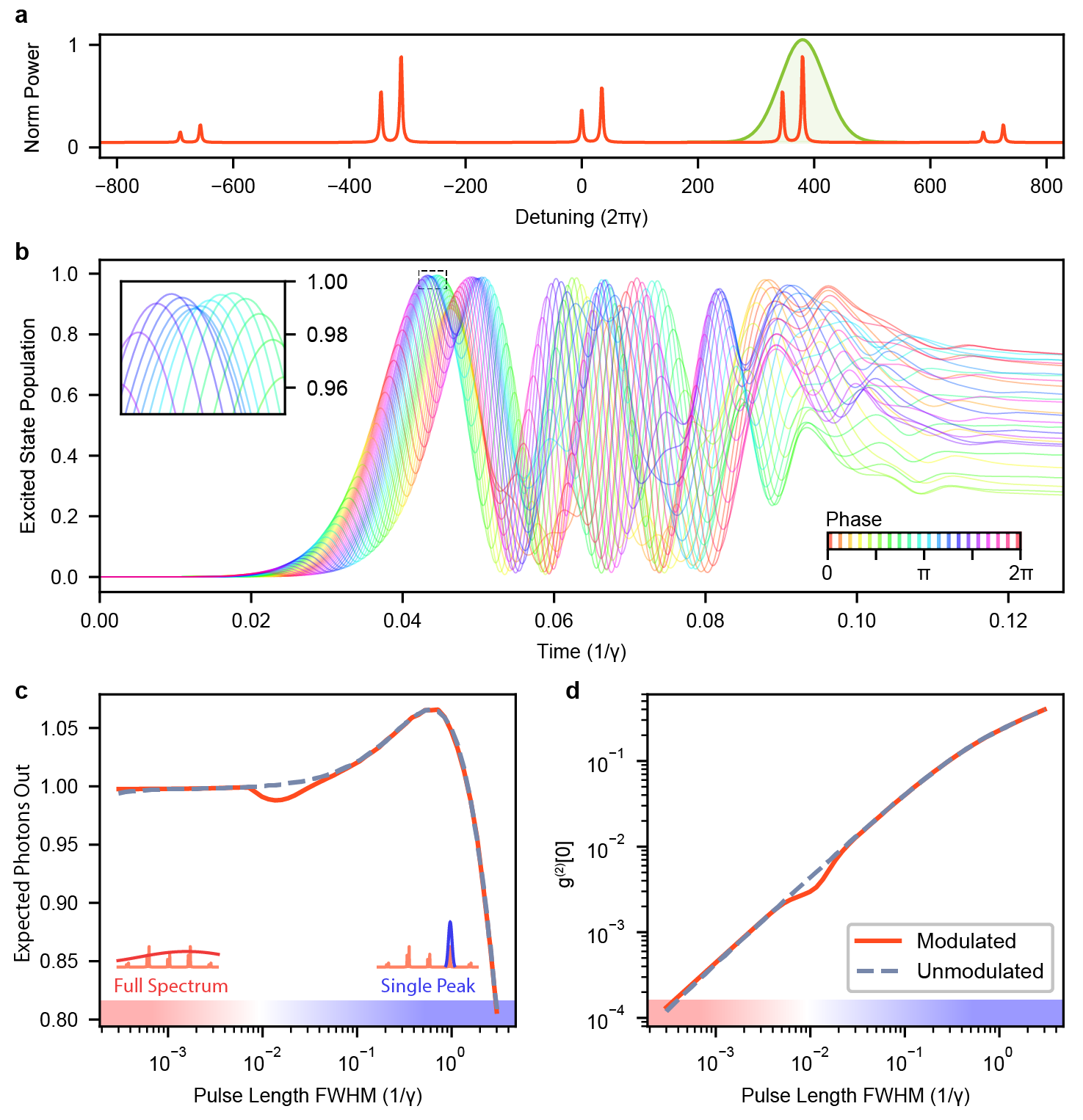}}
\end{figure*}

We have thus far considered a modulated TLS under continuous-wave optical excitation. However, it is also important to understand the behavior of the modulated TLS under pulsed optical excitation; this would determine whether a modulated emitter could be used as a deterministic source of shaped single photons. In a traditional TLS single-photon source, on-demand generation of highly-indistinguishable photons relies on coherent excitation with a short $\pi$-pulse, which prepares the system in the excited state with high fidelity, thus generating very nearly one-and-only-one photon per excitation\cite{FischerTwoPhoton2017}. The $\pi$-pulse must be resonant with the TLS, as a detuned pulse would induce off-resonant Rabi oscillations, which cannot efficiently transfer population.

To determine the feasibility of high-fidelity single-photon generation from a modulated quantum emitter, we investigate numerically the behavior of the modulated V\textsubscript{Si} under pulsed optical drive using experimental parameters corresponding to the high amplitude modulation in Supplementary Figure S7. We model the excited state population dynamics upon excitation with a short Gaussian optical pulse centered on one of the sidebands (Fig.~\ref{fig:pi_pulse_drive}a). We find that, unsurprisingly, the population dynamics are strongly dependent upon the phase of the periodic modulation $\Delta(t)$ relative to the arrival time of the optical pulse, as shown in Fig.~\ref{fig:pi_pulse_drive}b. Clearly, the simple $\pi$-pulse prescription used in an unmodulated TLS for complete population transfer from the ground to the excited state  does not apply. Instead, the phase and the pulse area are two free parameters that determine the single-photon generation fidelity (incidentally, both can be precisely controlled in experiment). For a range of experimentally-relevant pulse widths, we search the two-parameter space to compare the optimal single-photon generation fidelity of a modulated TLS to that of the unmodulated TLS: we evaluate the pulse-wise two-photon correlation function $g^{(2)}[0]$ as well as the expected number of emitted photons per pulse. Crucially, we find that the single-photon generation fidelity for a modulated TLS is very similar to that of the time-independent TLS\cite{UltraLowMultiPhoton2018}, reinforcing the potential utility of the modulated TLS as a deterministic single-photon source.

\begin{figure*}[h!]
\centering
\includegraphics[width=16cm]{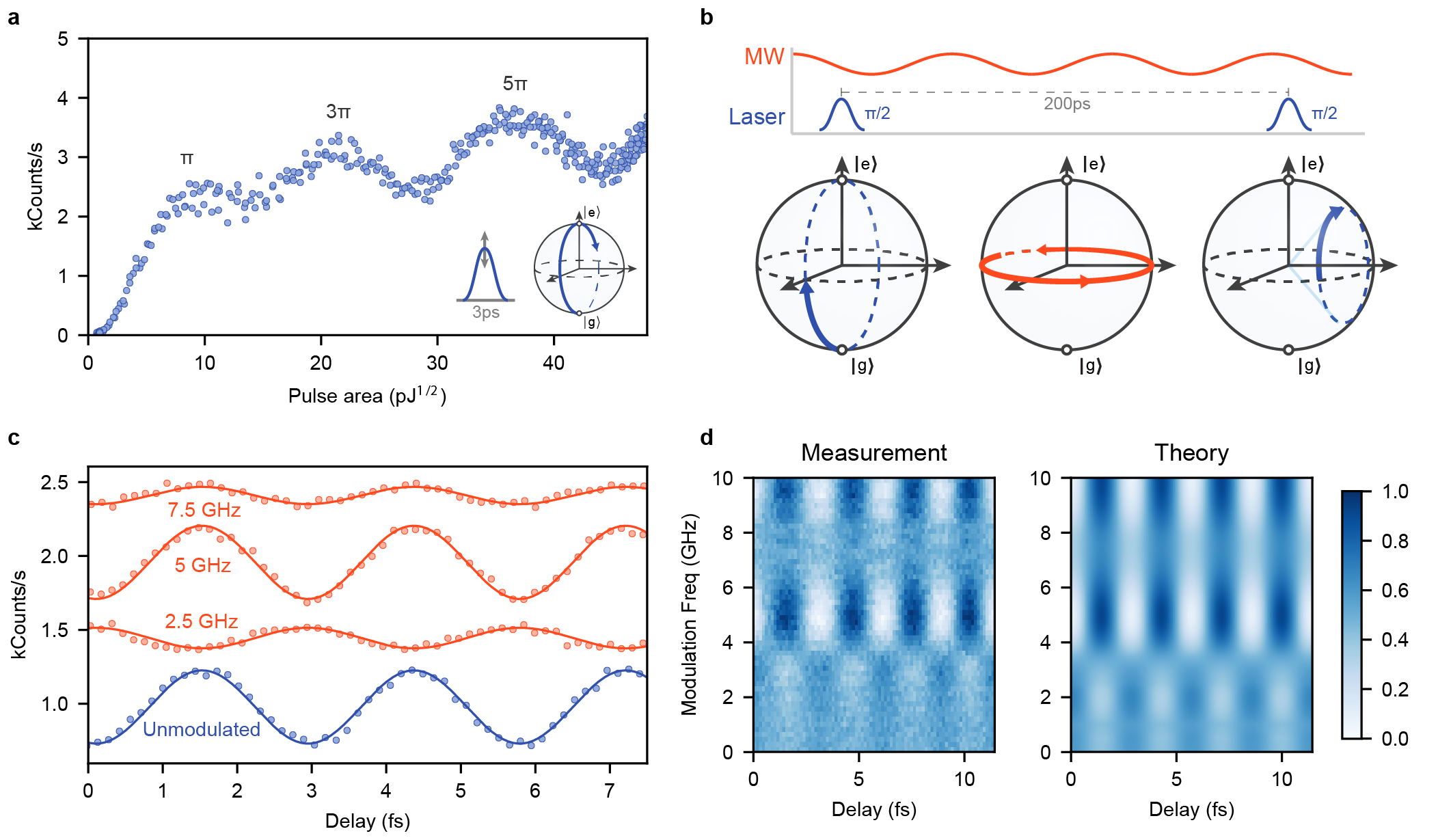}
\caption{\textbf{Figure 6 \textbar \hspace{0.1pt} Stark-modulated V\textsubscript{Si} interacting with short optical pulses.} \textbf{a.} Optical Rabi oscillations of a single unmodulated V\textsubscript{Si} excited by a 3~ps laser pulse.  \textbf{b.} To observe the effects of fast Stark modulation on the orbital state, we measure Ramsey interference by driving the V\textsubscript{Si} with two identical $\pi/2$ resonant pulses separated by a course delay of 200~ps. The Ramsey interference contrast will strongly depend on the modulation period relative to the interpulse delay. \textbf{c.} Observed Ramsey interference for various modulation frequencies, as well as for the unmodulated emitter. The data series are offset vertically for clarity. When the interpulse delay is an integer multiple of the modulation period, the observed interference is identical to that of the unmodulated V\textsubscript{Si}. \textbf{d.} Ramsey interference for modulation frequency swept from DC to 10~GHz. As predicted theoretically, full interference contrast is recovered at 5 and 10~GHz. \label{fig:ramsey}}
\end{figure*}

We proceed to investigate the interaction of a modulated V\textsubscript{Si} with short optical pulses. Using a resonant pulsed laser, we demonstrate fast control of the unmodulated V\textsubscript{Si} orbital state (Fig.~\ref{fig:ramsey}a). The high density of V\textsubscript{Si} in the sample induces a background fluorescence that limits the signal contrast. As the pulse bandwidth (3 ps) far exceeds the V\textsubscript{Si} modulation amplitude, modulation-induced orbital dynamics (shown in Fig.~\ref{fig:pi_pulse_drive}b) cannot be resolved with a single pulse. In order to observe signatures of Stark modulation in the orbital trajectories of the V\textsubscript{Si}, we perform a Ramsey interference experiment, where the V\textsubscript{Si} is manipulated by a pair of 3~ps optical $\pi/2$ pulses separated by 200~ps (Fig.~6b). We observe a strong dependence of the Ramsey interference amplitude on the modulation frequency (Fig.~6c). This effect is a consequence of the time-dependent Larmor precession experienced by the modulated V\textsubscript{Si} on the equator of the Bloch sphere. When the pulse delay is not a multiple of the modulation period, the accumulated interpulse precession depends on the phase of the microwave signal relative to the arrival of the first pulse. As we show in the Supplementary Information, the time-averaged Ramsey interference pattern is described by $1/2 + 1/2\cos(\omega t_{delay})J_0\bigg(2\frac{A}{\Omega}\sin(\frac{\Omega t_{delay}}{2})\bigg)$, where $t_{delay}$ is the time delay between the two $\pi/2$ pulses. We measure the Ramsey interference across different modulation frequencies and observe the recovery of the full Ramsey contrast at 5 and 10~GHz, in excellent agreement with the theoretical prediction (Fig.~\ref{fig:ramsey}d).

\section*{{\fontsize{11}{11}\textsf{\textbf{Conclusion\vspace{-1.2 ex}}}}}

In this Article, we have proposed and demonstrated spectral optimization of a quantum emitter and showed that a simple modulated two-level system can be used as a spectrally reconfigurable deterministic single-photon source. Using a scattering matrix formalism, we develop a rigorous model of this system, and use the model to engineer unconventional photon states. Using color centers modulated via the Stark shift, we experimentally demonstrate spectral shaping and study the interaction of a modulated optical transition with fast optical pulses. 

In light of the recent technological advances in SiC photonics\cite{Guidry2020Optical, song2019ultrahigh}, including the recent integration of single V\textsubscript{Si} into nanophotonic architectures\cite{4HSiCPhotonics2019}, our results suggest that the V\textsubscript{Si} is an excellent candidate for a scalable spectrally-reconfigurable single photon source. Furthermore, as this approach to spectral control of single photon emission requires only a rapidly modulated optical transition, it should be applicable to other solid-state defects modulated either via the Stark effect \cite{awschalom2018quantum, MiaoElectricallyCarbide, anderson2019electrical} or acoustically\cite{maity2020coherent, metcalfe2010resolved}.

As discussed in the Supplementary Information, the ability to rapidly chirp the emitter frequency enables the generation of spectrally-engineered chirped photons amenable to extreme temporal compression with additional dispersion correction.
Moreover, coherence-preserving rapid spectral modulation of an optical transition may have applications beyond spectral shaping. In atomic and superconducting qubit systems, frequency modulation has been proposed for simulating topological phase transitions\cite{silveri2017quantum}, overcoming dephasing \cite{li2013motional}, and implementing quantum gates using resolved sidebands\cite{beaudoin2012first}. TLS modulation based on the Stark effect is a flexible technique, compatible with integrated nanophotonic cavity systems which enhance atom-photon interactions\cite{Zhongeaan5959, Evans662}. Cavity integration would enable a solid-state implementation of the fast time-modulated Jaynes-Cummings system, which has received extensive theoretical investigation\cite{alsing1992dynamic, bagarello2015non}. In spin-based solid-state systems, where inhomogeneous broadening plagues the indistinguishability of photons emitted from different quantum nodes, optical transitions that are widely-tunable both statically and dynamically can open pathways toward scalable integrated quantum photonic systems. A unique application of high-fidelity fast control (which we explore numerically in the Supplementary Information) is the dynamic compensation of inhomogeneous broadening of an emitter ensemble via a single optimized microwave signal. In contrast with the traditional approach of statically tuning $N$ emitters on resonance using $N-1$ electrodes, this method requires just one set of electrodes and does not require spatially-separated emitters, making it uniquely suitable to improve photon-mediated spin-spin interactions in low-mode-volume nanophotonic cavities\cite{Evans662}.

\newpage

\section*{{\fontsize{11}{11}\textsf{\textbf{Methods\vspace{-1.2 ex}}}}}

\subsection*{{\fontsize{10}{10}\textsf{\textbf{Floquet spectrum optimization\vspace{-1.2 ex}}}}} When the emitter linewidth is much smaller than the modulation period, which is the case for our optimizations, the shape of the spectrum is dictated solely by the Fourier components of $\exp(-i\int_0^t\Delta(t') dt')$. Thus, we can define our desired Fourier series decomposition of our Floquet state and optimize the Fourier components of $\exp(-i\int_0^t\Delta(t') dt')$ to match those of the desired state. To do this we use the Broyden–Fletcher–Goldfarb–Shanno (BFGS) algorithm. Our optimization parameters are the real and imaginary parts of all Fourier series components of $\Delta(t)$ within the defined bandwidth, and our cost function is the mean squared error of the Fourier series components calculated with the discrete Fourier transform. By scaling the amplitude of $\Delta(t)$ by the modulation frequency $\Omega$, the spectral shape is conserved.

\subsection*{{\fontsize{10}{10}\textsf{\textbf{Sample Preparation\vspace{-1.2 ex}}}}}
The experiments were performed using a 100~\textmugreek m-thick 4H-\textsuperscript{28}Si\textsuperscript{12}C epilayer grown by chemical vapour deposition on a n-type (0001) 4H-SiC substrate. Color centers are generated via electron irradiation. In order to investigate whether the spectral stability of the V\textsubscript{Si} is influenced by the electron irradiation energy, one sample was irradiated with an average energy of 2~MeV (at QST, Japan) and another at an average energy of 23~MeV (at Stanford SLAC, USA), with a dose of $1\cdot10$\textsuperscript{13}~cm\textsuperscript{-2} and $5\cdot10$\textsuperscript{12}~cm\textsuperscript{-2}, respectively. Samples were annealed for 30 minutes at 300$^\circ$C after irradiation. Samples were diced and their edges were polished (DAG 810 from Disco Corp.). Then, 3~\textmugreek m were removed from the surface with reactive ion etching (using SF\textsubscript{6}), to minimize the presence of defects that arise from mechanical processing. Gold electrodes were patterned on the sample edge via e-beam lithography and liftoff. No difference in the properties of single V\textsubscript{Si} was observed between the two samples; however, as expected, the higher-energy irradiation produced a greater fraction of optically-active defects of unknown origin.

\subsection*{{\fontsize{10}{10}\textsf{\textbf{Experimental Setup\vspace{-1.2 ex}}}}} The measurements are performed in a closed-cycle cryostat (Montana Instruments) at a temperature of 5~K. The sample is mounted onto a custom-built circuit board with a microwave stripline optimized for high-frequency operation. The signal is delivered onto the sample with aluminum wirebonds. The cut-off frequency of the microwave setup was measured to be 10.5~GHz. Optical spectra of the V\textsubscript{Si} are measured via the PLE technique: by scanning a weak resonant laser (power at the objective lens ranging between 50 and 150~nW) across the transition, and detecting only the emission into the phonon side-band via a tunable long-pass filter (Semrock). Two-photon coincidences are recorded with timing electronics with a 10~ps resolution (Swabian Instruments). To control the charge state of the emitter, a 1~\textmugreek s above-resonant (740 nm) repump pulse is applied at a 1~kHz repetition rate. For pulsed measurements, a picosecond Ti:Sapphire laser (Spectra Physics) with a home-built pulse delay stage and and EOM-based pulse picker are used. For DC Stark tuning characterization, voltage is applied to the gold electrodes via a programmable voltage source (Keithley). Single-frequency microwave drive is delivered via a continuous-wave signal generator with 3.3~GHz bandwidth (Rhode-Shwartz). Engineered multi-frequency microwave drives are generated by an arbitrary waveform generator (Keysight) with amplification (MiniCircuits). A diagram of the optical and electronic experimental setup is shown in Supplementary Figures S1 and S2.

\section*{{\fontsize{11}{11}\textsf{\textbf{Data availability\vspace{-1.2 ex}}}}}
All data relevant to the current study are available from the corresponding author on request. {\color{white}\cite{xu2015input, trivedi2019photon, gardiner1985input}} 

\section*{{\fontsize{11}{11}\textsf{\textbf{Code availability\vspace{-1.2 ex}}}}}
The code used for scattering matrix simulations can be accessed at
\begin{verbatim}
    https://github.com/rahultrivedi1995/oqs_scattering
\end{verbatim} 

\section*{{\fontsize{11}{11}\textsf{\textbf{References\vspace{-1.2 ex}}}}}

\begingroup
\renewcommand{\section}[2]{}%
\bibliography{sample}

\endgroup


\section*{{\fontsize{11}{11}\textsf{\textbf{Acknowledgements\vspace{-1.2 ex}}}}}
This research is funded in part by the U.S. Department of Energy, Office of Science, under Awards DE-SC0019174 and DE-Ac02-76SF00515; and the National Science Foundation under award 1839056. Part of this work was performed at the Stanford Nanofabrication Facility (SNF) and the Stanford Nano Shared Facilities (SNSF), supported by the National Science Foundation under award ECCS-1542152. 
D.L. acknowledges support from the Fong Stanford Graduate Fellowship (SGF) and the National Defense Science and Engineering Graduate Fellowship. 
A.D.W. acknowledges support from the Herb and Jane Dwight SGF. 
M.A.G. acknowledges support from the Albion Hewlett SGF and the NSF Graduate Research Fellowship. 
R.T. acknowledges funding from Kailath Graduate Fellowship.
N.T.S. acknowledges funding by the Swedish Research Council (Vetenskapsradet VR 2016-04068). 
J.U.H. acknowledges funding by the Swedish Energy Agency (43611-1). N.T.S. and J.U.H. acknowledge funding by the EU H2020 project QuanTELCO (862721) and the Knut and Alice Wallenberg Foundation (KAW 2018.0071).
T.O. acknowledges support from grants JSPS KAKENHI 17H01056 and 18H03770.
J.W. acknowledges support by the European Research Council (ERC) grant SMel, the European Commission Marie Curie ETN ``QuSCo'' (GA No 765267), the Max Planck Society, the Humboldt Foundation, the German Science Foundation (SPP 1601), and the EU-FET Flagship on Quantum Technologies through the project ASTERIQS. 
J.W. and F.K. acknowledge the EU-FET Flagship on Quantum Technologies through the project QIA.
C.D. acknowledges support from the Andreas Bechtolsheim SGF and the Microsoft Research PhD Fellowship. 

\section*{{\fontsize{11}{11}\textsf{\textbf{Author Contributions\vspace{-1.2 ex}}}}}
DML, ADW, MAG, JV  conceived the experiment.
MAG, DML, ADW built the experimental setup.
DML, ADW, MAG conducted the experiment.
RT, ADW, DML, OOS conducted the theoretical analysis.
ADW, RT performed microwave engineering optimization.
NM, CB, CD, FK, JW assisted with experimental setup and material characterization.
JUH, NTS designed and performed SiC growth.
TO, PKV, MHN, EAN performed the electron irradiation.
SS, JPWM provided experimental and theoretical guidance.
JV supervised the project.
All authors discussed the results and contributed to the final version of the manuscript.
 \section*{{\fontsize{11}{11}\textsf{\textbf{Correspondence\vspace{-1.2 ex}}}}} Correspondence and requests for materials
should be addressed to J.V.~(email: jela@stanford.edu).

\end{document}